\documentclass[aps,prd,showpacs,showkeys,superscriptaddress,twocolumn]{revtex4-1}
\usepackage[colorlinks,linkcolor=blue,anchorcolor=blue,citecolor=blue,urlcolor=blue,breaklinks=true]{hyperref}
\usepackage{graphicx}
\usepackage{amsmath}
\usepackage{amssymb}
\usepackage{slashed}
\usepackage{latexsym}
\usepackage{epsfig}
\usepackage{amsbsy}
\usepackage{array}
\usepackage{changes}
\usepackage{amssymb}
\usepackage{setspace}
\usepackage{bm}
\usepackage{lipsum}
\usepackage{mathrsfs}
\usepackage{float}
\usepackage{color}
\usepackage{graphicx}
\usepackage{subfigure}
\usepackage[T1]{fontenc}
\usepackage{mathptmx}
\DeclareMathAlphabet{\mathcal}{OMS}{cmsy}{m}{n}
\DeclareSymbolFont{largesymbols}{OMX}{cmex}{m}{n}

\RequirePackage[normalem]{ulem} 
\RequirePackage{color}\definecolor{RED}{rgb}{1,0,0}\definecolor{BLUE}{rgb}{0,0,1} 
\providecommand{\DIFaddbegin}{} 
\providecommand{\DIFaddend}{} 
\providecommand{\DIFdelbegin}{} 
\providecommand{\DIFdelend}{} 
\providecommand{\DIFaddbeginFL}{} 
\providecommand{\DIFaddendFL}{} 
\providecommand{\DIFdelbeginFL}{} 
\providecommand{\DIFdelendFL}{} 
\newcommand{\DIFscaledelfig}{0.5}
\RequirePackage{settobox} 
\RequirePackage{letltxmacro} 
\newsavebox{\DIFdelgraphicsbox} 
\newlength{\DIFdelgraphicswidth} 
\newlength{\DIFdelgraphicsheight} 
\LetLtxMacro{\DIFOincludegraphics}{\includegraphics} 
\newcommand{\DIFaddincludegraphics}[2][]{{\color{blue}\fbox{\DIFOincludegraphics[#1]{#2}}}} 
\newcommand{\DIFdelincludegraphics}[2][]{
\sbox{\DIFdelgraphicsbox}{\DIFOincludegraphics[#1]{#2}}
\settoboxwidth{\DIFdelgraphicswidth}{\DIFdelgraphicsbox} 
\settoboxtotalheight{\DIFdelgraphicsheight}{\DIFdelgraphicsbox} 
\scalebox{\DIFscaledelfig}{
\parbox[b]{\DIFdelgraphicswidth}{\usebox{\DIFdelgraphicsbox}\\[-\baselineskip] \rule{\DIFdelgraphicswidth}{0em}}\llap{\resizebox{\DIFdelgraphicswidth}{\DIFdelgraphicsheight}{
\setlength{\unitlength}{\DIFdelgraphicswidth}
\begin{picture}(1,1)
\thicklines\linethickness{2pt} 
{\color[rgb]{1,0,0}\put(0,0){\framebox(1,1){}}}
{\color[rgb]{1,0,0}\put(0,0){\line( 1,1){1}}}
{\color[rgb]{1,0,0}\put(0,1){\line(1,-1){1}}}
\end{picture}
}\hspace*{3pt}}} 
} 
\LetLtxMacro{\DIFOaddbegin}{\DIFaddbegin} 
\LetLtxMacro{\DIFOaddend}{\DIFaddend} 
\LetLtxMacro{\DIFOdelbegin}{\DIFdelbegin} 
\LetLtxMacro{\DIFOdelend}{\DIFdelend} 
\DeclareRobustCommand{\DIFaddbegin}{\DIFOaddbegin \let\includegraphics\DIFaddincludegraphics} 
\DeclareRobustCommand{\DIFaddend}{\DIFOaddend \let\includegraphics\DIFOincludegraphics} 
\DeclareRobustCommand{\DIFdelbegin}{\DIFOdelbegin \let\includegraphics\DIFdelincludegraphics} 
\DeclareRobustCommand{\DIFdelend}{\DIFOaddend \let\includegraphics\DIFOincludegraphics} 
\LetLtxMacro{\DIFOaddbeginFL}{\DIFaddbeginFL} 
\LetLtxMacro{\DIFOaddendFL}{\DIFaddendFL} 
\LetLtxMacro{\DIFOdelbeginFL}{\DIFdelbeginFL} 
\LetLtxMacro{\DIFOdelendFL}{\DIFdelendFL} 
\DeclareRobustCommand{\DIFaddbeginFL}{\DIFOaddbeginFL \let\includegraphics\DIFaddincludegraphics} 
\DeclareRobustCommand{\DIFaddendFL}{\DIFOaddendFL \let\includegraphics\DIFOincludegraphics} 
\DeclareRobustCommand{\DIFdelbeginFL}{\DIFOdelbeginFL \let\includegraphics\DIFdelincludegraphics} 
\DeclareRobustCommand{\DIFdelendFL}{\DIFOaddendFL \let\includegraphics\DIFOincludegraphics} 
\RequirePackage{listings} 
\RequirePackage{color} 
\lstdefinelanguage{DIFcode}{ 
  moredelim=[il][\color{red}\sout]{\%DIF\ <\ }, 
  moredelim=[il][\color{blue}\uwave]{\%DIF\ >\ } 
} 
\lstdefinestyle{DIFverbatimstyle}{ 
	language=DIFcode, 
	basicstyle=\ttfamily, 
	columns=fullflexible, 
	keepspaces=true 
} 
\lstnewenvironment{DIFverbatim}{\lstset{style=DIFverbatimstyle}}{} 
\lstnewenvironment{DIFverbatim*}{\lstset{style=DIFverbatimstyle,showspaces=true}}{} 

\begin{document}
\author{Zheng Zhang}
\email{jozhzhang@163.com}
\affiliation{Department of Physics, Nanjing University, Nanjing 210093, China}
\author{Chao Shi}
\email{cshi@nuaa.edu.cn}
\affiliation{Department of Nuclear Science and Technology,
Nanjing University of Aeronautics and Astronautics, Nanjing 210016, China}
\author{Xiao-Tao He}
\email{hext@nuaa.edu.cn}
\affiliation{Department of Nuclear Science and Technology,
Nanjing University of Aeronautics and Astronautics, Nanjing 210016, China}
\author{Xiaofeng Luo}
\email{xfluo@ccnu.edu.cn}
\affiliation{Key Laboratory of Quark and Lepton Physics (MOE) and Institute of Particle Physics, Central China Normal University, Wuhan 430079, China}
\author{Hong-Shi Zong}
\email{zonghs@nju.edu.cn}
\affiliation{Department of Physics, Nanjing University, Nanjing 210093, China}
\affiliation{Nanjing Proton Source Research and Design Center, Nanjing 210093, China}
\affiliation{Department of Physics, Anhui Normal University, Wuhu, Anhui 241000, China }
\date{\today}

\title{Chiral phase transition inside a rotating cylinder within the Nambu--Jona-Lasinio model}

\begin{abstract}
{We study the chiral phase transition inside a rotating cylinder within the framework of the Nambu--Jona-Lasinio model. A spectral boundary condition is imposed to avoid faster than light. We investigate how the geometry of the cylinder and rotation influence the chiral phase transition at finite temperature and chemical potential. The inhomogeneous effects caused by the finite size and rotation are also taken into account. It is found that finite size will reduce the chiral transition temperature and raises the chiral transition chemical potential, while the rotation reduces both the chiral transition temperature and chemical potential. In addition, we discuss the implications of our results in heavy-ion collisions and equation of states of neutron star. }
\bigskip


\end{abstract}

\maketitle


\maketitle

\section{Introduction}
The properties of rotating strongly interacting matter have got a lot of attention in recent years, which is mainly driven by the noncentral heavy ion collisions (HICs). In these experiments, the quark-gluon plasma produced carry large angular momentum and have high vorticity \cite{vorticalfluid}. Another place where rotation may play an important role is the pulsars, which can rotate rapidly \cite{starreview}. Rotating can induce some interesting transport phenomenon in strongly interacting matter like the chiral vortical effect \cite{chiralv1,chiralv2,chiralv3}, which is analogy to the chiral magnetic effect \cite{chiralv3, Yinke}.  The phase transition can also be influenced by rotation. In this paper, we concentrate on the effect of rotation on the chiral phase transition.

There are some studies about the effect of rotation on chiral phase transition.  For example, the phase transition of Nambu--Jona-Lasinio (NJL) type models under rotation was discussed in Refs. \cite{YinJiang,bound1,cylinder}. The interplay between rotation and magnetic field was discussed in Ref. \cite{FukushimaH}.  The properties of rotating QGP systems were also studied with holographic method \cite{holo3}. However, there are some complexities in discussing the effect of rotation on the chiral phase transition. First, a rigidly rotating system should be bounded in the directions perpendicular to the rotating axis, otherwise some region of the system will exceed the speed of light.  This implies that a rotating system must also be a finite size system (at least in the directions perpendicular to the rotating axis). When the angular velocity is very high, the system must be very small. Then the finite size effects must be considered because it also influences the chiral phase transition, see a review \cite{reviewK}. Second, the rotation and finite size will induce inhomogeneous distributions, for example, as we shall show, the condensate varies with coordinates. Considering both the two points, we will study the effect of rotation and finite size on the chiral phase transition with the NJL model.

To bound the system inside a finite region, one should impose a boundary condition. The selection of boundary condition is very important because the results depend on the boundary conditions \cite{reviewK}. In our previous studies \cite{zzhang1,zzhang2}, we adopted the MIT boundary condition \cite{MIT1}, which ensures the normal component of particle current  to vanish on the surface. However, this boundary condition breaks the chiral symmetry explicitly \cite{cylinder} and thus seems not very realistic. (The treatment in Refs. \cite{zzhang1,zzhang2} do not reflect the fact that the chiral symmetry is explicitly broken by the MIT boundary condition, so the results obtained need futher discussion.)  In this paper, we adopt the spectral boundary condition \cite{spectral,rcylinder}, which preserves the self-adjointness of the Hamiltonian and preserves the chiral symmetry for a cylindrical boundary. This boundary condition has also been adopted in Ref. \cite{bound1} to investigate the rotating NJL model, but only the case at zero temperature and zero chemical potential are studied. 

Before discussing the details of calculation, we would like to make some comments on the effects of rotation and finite size.  In previous studies, the rotation is found to suppress the chiral condensate \cite{, YinJiang,FukushimaH} at finite temperature, which can even restore the chiral symmetry at zero temperature and zero chemical potential.  However, both Ref. \cite{YinJiang} and Ref. \cite{FukushimaH} treat the rotating system to be unbounded. By constraining a rotating system inside a region not exceeding the speed of light, it is shown that at zero temperature and zero chemical potential, rotation has no effect on the chiral condensate \cite{cylinder, bound1, zzhang2}, at least in the NJL model. This conclusion can find its root in the spectrum of free fermions in a rotating frame, which determines whether the rotating vacuum and nonrotating vacuum are identical \cite{rcylinder}.  It is suggested by Refs. \cite{rcylinder, zzhang3} that the two vacuum should be identical.  And in this case, one can show the rotation have no effect at zero temperature and zero chemical potential. We will demonstrate this relation with more detail below. This conclusion also makes the phase diagram in the temperature-rotation parameter space suggested in \cite{YinJiang} questionable, although the rotation can suppress the condensate is still qualitatively right.   
 In some aspects, a rotation is similar with a chemical potential (or finite density) \cite{YinJiang, FukushimaH}, so the rotation tends to decrease the condensate and thus lower the critical temperature and chemical potential. As we will see, the Hamiltonian of a free particle in rotation frame is shifted as $H-\boldsymbol{\Omega}\cdot \mathrm{\boldsymbol{J}}$, where $\boldsymbol{\Omega}$ is the angular velocity and $\mathrm{\boldsymbol{J}}$ is the angular momentum.  Thus the spectrum of free fermion will be shifted as $E-\Omega j_z$, where $j_z$ is the $z$ component of angular momentum (we set the angular velocity in the $z$ direction).  $\Omega j_z$ is analogy to a effective chemical potential. But we would like to point out, this effective chemical potential depends on $j_z$, and when calculating thermal expectation values, one should sum over all states with different $j_z$, which makes rotation different from a chemical potential in certain aspects. As for finite size effects, previous studies show that it also suppresses the chiral condensate \cite{qingwu,Xia}. This is easy to understand since spontaneous breaking can only happen in infinite systems in principle \cite{Weinberg2}. Finite size is often compared to finite temperature, since the temperature is the inverse of the temporal system length scale. So, it is expected that a small size leads to similar effects with a high temperature. For example, small size will decrease the effective mass as high temperature does. Thus small size will decrease the chiral transition temperature \cite{qingwu,Palhares2011,Xia}. However, the analogy between finite size and temperature also has its own limit. Sometimes, finite size has different effects with high temperature. It is found that finite size can raise the chiral transition chemical potential \cite{Palhares2011,MRE4,DS2,DS4}, while high temperature decreases the transition chemical potential. Our results in this paper are consistent with Refs. \cite{Palhares2011,MRE4,DS2,DS4}. 

The novelty of this paper is that it considers the finite size, rotation and inhomogeneous effects at the same time. Not only the chiral transition at finite temperature, but also the chiral transition at finite chemical potential is studied. Also, the boundary condition imposed to constrain the system inside a region can avoid possible unphysical results due to faster than light.

The structure of this paper is as follows:  In Sec. \ref{II}, we briefly introduce the mode solutions of free fermions and the chiral condensate in rotating coordinates with cylindrical spectral boundary condition. In Sec. \ref{III}, we give the NJL model inside a rotating cylinder and discuss the chiral phase transition at finite temperature. The influences of rotation and finite size are discussed and the inhomogeneous effects are considered. We also discussed the chiral phase transition at a finite chemical potential. A summary and discussion is given in Sec. \ref{IV}, where we discuss the implications of our results in heavy-ion collisions and the equation of states of neutron stars.

\section{Mode solutions and condensate} \label{II}
Let us briefly introduce the mode solutions of  free fermions inside a cylinder with spectral boundary condition. Details can be found in Ref. \cite{rcylinder}, while here we only recapitulate the main results. In the paper, we adopt the units $\hbar=c=k_B=1$.

We set the rotation axis to be the $z$ axis, the metric in rotating coordinates with angular velocity $\Omega$ is
\begin{equation}
g_{\mu \nu}=\left(\begin{array}{cccc}
{1-\left(x^{2}+y^{2}\right) \Omega^{2}} & {y \Omega} & {-x \Omega} & {0} \\
{y \Omega} & {-1} & {0} & {0} \\
{-x \Omega} & {0} & {-1} & {0} \\
{0} & {0} & {0} & {-1}
\end{array}\right).
\end{equation}
We use $t,x,y,z$ to represent the rotating coordinates and $\hat{t},\hat{x},\hat{y},\hat{z}$ to represent the static Cartesian coordinates.
The Dirac equation with a general metric is 
\begin{equation}
[i\gamma^\mu(\partial_\mu+\Gamma^\mu)-M]\psi=0,
\end{equation}
where 
\begin{equation}
\begin{aligned}
&\Gamma_\mu=-\frac{i}{4}\omega_{\mu \hat{i}\hat{j}}\sigma^{\hat{i}\hat{j}},\\
&\omega_{\mu \hat{i}\hat{j}}=g_{\alpha\beta}e^\alpha_{\hat{i}}(\partial_\mu e^{\beta}_{\hat{j}}+\Gamma^{\beta}_{\nu\mu} e^\mu_{\hat{j}}),\\
&\sigma^{\hat{i}\hat{j}}=\frac{i}{2}[\gamma^{\hat{i}},\gamma^{\hat{j}}],
\end{aligned}
\end{equation}
with the Christoffel connection, $\Gamma^\lambda_{\mu\nu}=\frac{1}{2}g^{\lambda\sigma}(g_{\sigma\nu,\mu}+g_{\mu\sigma.\nu}-g_{\mu\nu,\sigma})$, and the gamma matrix in curved space-time, $\gamma^\mu=e^\mu_{\hat{i}}\gamma^{\hat{i}}$. The vierbein $e^\mu_{\hat{i}}$ connects the general coordinate with the Cartesian coordinate in the rest frame, $x^\mu=e^\mu_{\hat{i}}x^{\hat{i}}$. Then the Dirac equation in rotating coordinates can be reduced to \cite{cylinder}
\begin{equation}
\label{Dirac}
\left[\gamma^{\hat{t}}\left(i \partial_{t}+\Omega J_{z}\right)+i \gamma^{\hat{x}} \partial_{x}+i \gamma^{\hat{y}} \partial_{y}+i \gamma^{\hat{z}} \partial_{z}-M\right] \psi=0.
\end{equation}
To solve this equation, one can suppose the particle mode solution to be
\begin{equation}
\psi(x)=u(x)e^{-i\widetilde{E}t}.
\end{equation}
Then we have the time-independent equation:
\begin{equation} \label{eq1}
\widetilde{H}u(x)=\widetilde{E}u(x),
\end{equation}
where 
\begin{equation}
\widetilde{H}=-i\gamma^{\hat{0}}\gamma^{\hat{i}}\partial_i+\gamma^{\hat{0}}M-\Omega J_z=H-\Omega J_z.
\end{equation}
$H$ has the same form with the free Hamiltonian in the rest frame. To solve Eq. (\ref{eq1}), we find a set of commutating operators $\{H,P_z, J_z, W_0\}$, where $W_0$ is the helicity operator:
\begin{equation}W_{0}=\left(\begin{array}{ll}
h & 0 \\
0 & h
\end{array}\right), \quad h=\frac{\boldsymbol{\sigma} \cdot \boldsymbol{P}}{2 p},\end{equation}
where $\boldsymbol{\sigma}$ are the Pauli matrices. We can lable a eigenstate $u_j(x)$ by its eigenvalues:
\begin{equation}
j=(E_j, k_j, m_j,\lambda_j).
\end{equation}
where $m_j=0,\pm 1,\pm2,...$, and $\lambda_j=\pm\frac{1}{2}$.  Here we follow the convention used in \cite{rcylinder} that $m_j+\frac{1}{2}$ rather than $m_j$ to be the eigenvalue of $J_z$. The corotating energy $\widetilde{E_j}$ is related to the Minkowski energy $E_j$ by :
\begin{equation}
\widetilde{E_j}=E_j-\Omega (m_j+\frac{1}{2}).
\end{equation}
The mode solution to Eq. (\ref{eq1}) is given as \cite{rcylinder}: 
\begin{equation}u_{j}(r, \varphi,z)=\frac{1}{\sqrt{2}}\left(\begin{array}{c}
\mathrm{E}_{+} \phi_{j} \\
\frac{2\lambda E}{|E|} \mathrm{E}_{-} \phi_{j}
\end{array}\right)\frac{e^{ikz}}{2\pi},\end{equation}
\begin{equation}\phi_{j}(r, \varphi)=\frac{1}{\sqrt{2}}\left(\begin{array}{c}
\mathrm{p}_{\lambda} e^{i m \varphi} J_{m}(q r) \\
2 i \lambda \mathrm{p}_{-\lambda} e^{i(m+1) \varphi} J_{m+1}(q r)
\end{array}\right),\end{equation}
where 
\begin{equation}
\mathrm{E}_{\pm}=\sqrt{1\pm \frac{M}{E}},\ \ \mathrm{p}_{\pm}\equiv\mathrm{p}_{\pm 1/2}=\sqrt{1\pm \frac{k}{p}},
\end{equation}
$E=\pm \sqrt{p^2+M^2}$ and $p=\sqrt{q^2+k^2}$. $J_m$ is the $m$th ordered Bessel function.
We can write the particle mode solution to Eq. (\ref{Dirac}) as
\begin{equation}
U_j(r,\varphi, z,t)=u_j(r, \varphi,z)e^{-i\widetilde{E_j}t}.
\end{equation}
The antiparticle mode solution $V_j(x)$ can be obtained by the charge conjugate:
\begin{equation}
V_j(x)=i\gamma^{\hat{2}}U_j^{*}(x).
\end{equation}
These are solutions in unbounded space-time. To find the solution bounded in a cylindrical boundary, we impose the spectral boundary condition \cite{rcylinder}:
\begin{equation}\begin{array}{ll}
\psi_{m+\frac{1}{2}}^{1}|_{r=R}=\psi_{m+\frac{1}{2}}^{3}|_{r=R}=0, & \text { for } m+\frac{1}{2}>0 \\
\psi_{m+\frac{1}{2}}^{2}|_{r=R}=\psi_{m+\frac{1}{2}}^{4}|_{r=R}=0, & \text { for } m+\frac{1}{2}<0,
\end{array}\end{equation}
where the upper index 1,2,3,4 means the 1,2,3,4 component of the spinor and the lower index $m+1/2$ labels the angular momentum of the mode solution.  $R$ is the radius of the cylinder. With this boundary condition, the transverse momentum $q$ is discrtized as 
\begin{equation}q_{m, \ell} R=\left\{\begin{array}{ll}
\xi_{m, \ell} & m+\frac{1}{2}>0 \\
\xi_{-m-1, \ell} & m+\frac{1}{2}<0,
\end{array}\right.\end{equation}
where $\xi_{m,\ell}$ is the $\ell$th nonzero root of the Bessel function $J_m$. Now, the eigenstates should be labeled by:
\begin{equation}
j=(E_j, k_j, m_j, \lambda_j, \ell_j).
\end{equation}
And the mode solutions with spectral boundary condition can be normalized as
\begin{equation}
U^{\mathrm{sp}}_j=C^{\mathrm{sp}}_j U_j,
\end{equation}
where the coefficient $C^{\mathrm{sp}}_j$ is given as \cite{rcylinder}
\begin{equation}
C^{\mathrm{sp}}_j=\mathcal{C}_{E k m \ell}^{\lambda, \mathrm{sp}}=\mathcal{C}_{E, k,-m-1, \ell}^{\lambda, \mathrm{sp}}=\frac{\sqrt{2}}{R\left|J_{m+1}\left(\xi_{m, \ell}\right)\right|}
\end{equation}
for positive $m$. 

After we get the free fermions spectrum and the mode solutions, we can get the expression of the free field fermions condensate $\langle\overline{\psi}\psi\rangle$. Here we follow the approach in Ref. \cite{zzhang3}, while the main results of $\langle\overline{\psi}\psi\rangle$ has been obtained in Ref. \cite{rcylinder}.

To obtain the expression of $\langle\overline{\psi}\psi \rangle$, we first expand the field operator $\psi$ by the mode solutions,
\begin{equation}\psi=\sum_{j} \theta\left(E_{j}\right)\left[U_{j}^{\mathrm{sp}} b_{j}+V_{j}^{\mathrm{sp}} \mathrm{d}_{j}^{\dagger}\right],\end{equation}
where $\theta(E_j)$ is the step function and the sum over $j$ is the abbreviation for:
\begin{equation}
\sum_{j} \equiv \sum_{\lambda_{j}=\pm 1 / 2} \sum_{m_{j}=-\infty}^{\infty} \sum_{\ell_{j}=1}^{\infty} \int_{-\infty}^{\infty} d k_{j} \sum_{E_{j}=\pm\left|E_{j}\right|}.
\end{equation}
We also expand $\overline{\psi}$ by mode solutions,
\begin{equation}\overline{\psi}=\sum_{j} \theta\left(E_{j}\right)\left[\overline{U}_{j}^{\mathrm{sp}} b_{j}^\dagger+\overline{V}_{j}^{\mathrm{sp}} \mathrm{d}_{j}\right].\end{equation}
Multiple $\overline{\psi}$ and $\psi$ and take the ensemble average. Use the expression of mode solutions and the following distributions:
\begin{equation}
\begin{aligned}
&\langle b_j^{\dagger}b_{j'}\rangle=\frac{1}{e^{\beta(\widetilde{E_j}-\mu)}+1}\delta(j,j'),\\
&\langle d_jd_{j'}^{\dagger}\rangle=1-\langle d_{j'}^{\dagger} d_j\rangle
=(1-\frac{1}{e^{\beta(\widetilde{E_j}+\mu)}+1})\delta(j,j'),\\
&\langle b_j^{\dagger}d_{j'}^{\dagger}\rangle=\langle d_jb_{j'}\rangle =0.
\end{aligned}
\end{equation}
one then gets the condensate
\begin{equation}\label{conden}
\langle \overline{\psi} \psi\rangle=-\sum_{m=0}^{\infty} \sum_{\ell=1}^{\infty} \int_{0}^{\infty} \frac{M d k}{E \pi^{2} R^{2}} \frac{w(\widetilde{E})+w(\overline{E})}{J_{m+1}^{2}(q R)} J_{m}^{+}(q r),
\end{equation}
where 
\begin{equation}
J_m^{+}(x)=J_m^2(x)+J_{m+1}^2(x),
\end{equation}
\begin{equation}
w(E)=1-\frac{1}{1+e^{\beta(E-\mu)}}-\frac{1}{1+e^{\beta(E+\mu)}},
\end{equation}
$E=\sqrt{q^2+k^2+M^2}$ and $\overline{E}=E+\Omega(m+\frac{1}{2})$.
Here we note the above result can be applied to both finite temperature and finite chemical potential.  Due to the cylindrical symmetry, the condensate only depend on coordinate $r$.

Now, we can explain why rotation has no effect on chiral condensate $\langle\overline{\psi}\psi\rangle$ at zero temperature and zero chemical potential. The dependence of  $\langle\overline{\psi}\psi\rangle$ on angular velocity $\Omega$ only through $w(\widetilde{E})$ and $w(\overline{E})$. At zero temperature and zero chemical potential,
\begin{equation}
w(\widetilde{E})=1-\theta(-\widetilde{E}),\ \  w(\overline{E})=1-\theta(-\overline{E}).
\end{equation}
Since $\overline{E}>0$, one only need to consider the sign of $\widetilde{E}$. It is shown by Ref. \cite{rcylinder} that if the system is enclosed in a cylindrical boundary with $\Omega R<1$, there must be $E\widetilde{E}>0$. Thus, the condensate is independent of $\Omega$ at zero temperature and zero chemical potential. Meanwhile, $E\widetilde{E}>0$ makes the rotating vacuum and the nonrotating vacuum equivalent \cite{rcylinder}, which justifies our conclusion in the Introduction. This conclusion, is not only applicable for free fermion field, but also the NJL model inside a cylinder in a mean-field sense, since the expression of $\langle \overline{\psi}\psi \rangle$ in the NJL model has the same form with Eq. (\ref{conden}).

\section{Chiral phase transition inside a Rotating cylinder}\label{III}
Now, we can modify the NJL model to study the chiral phase transition in a rotating cylinder.
The Lagrangian of the two-flavor NJL model is  \cite{NJLreview}
\begin{equation}
\mathscr{L}=\overline{\psi}(i\gamma^\mu\partial_\mu-m_0)\psi+G[(\overline{\psi}\psi)^2+(\overline{\psi}i\gamma^5\tau\psi)^2],
\end{equation}
where $m_0$ is the current quark mass, and $G$ is the effective coupling. In the mean field approximation (only consider the Hartree term), the gap equation is given as 
\begin{equation}\label{gap}
M=m_0-2GN_cN_f\langle\overline{\psi}\psi\rangle.
\end{equation}
In Eq. (\ref{gap}), $\psi$ is in the spinor space. $N_c=3$ and $N_f=2$ is the number of colors and flavors respectively. In infinite space, $\langle\overline{\psi}\psi\rangle$ is homogeneous, and we can treat $\langle\overline{\psi}\psi\rangle$ as a free field chiral condensate with an effective mass $M$. However, in a finite size system, the condensate is inhomogeneous in general, and a rotation will also induce inhomogeneous condensate. Thus, $\langle\overline{\psi}\psi\rangle$ depends on coordinates. This dependence makes it is hard to solve Eq. (\ref{gap}) self-consistently. Here, we adopt the local density approximation, which has been adopted in Refs. \cite{YinJiang, bound1}. This approximation uses the free field condensate at point $x$ to replace the true condensate at point $x$ in Eq. (\ref{gap}). For the NJL model in a rotating cylinder, that is,
\begin{equation}\label{gap2}
M=m_0+2GN_cN_f\sum_{m=0}^{\infty} \sum_{\ell=1}^{\infty} \int_{0}^{\infty} \frac{M d k}{E \pi^{2} R^{2}} \frac{w(\widetilde{E})+w(\overline{E})}{J_{m+1}^{2}(q R)} J_{m}^{+}(q r).
\end{equation}
Here we note that the local density approximation is only valid when the effective mass varies slowly ($|\partial_r m| \ll m^2$) \cite{bound1}, so only the part where the effective mass varies slowly is reliable.
Now we can solve equation (\ref{gap2}) at every $r$ self-consistently to investigate the chiral phase transition. To do this, one still need to perform regularization, since the right hand of Eq. (\ref{gap2}) is divergent. Here we use the proper-time regularization, which take into account all the modes. The key equation of this regularization is the replacement:
\begin{equation}
\frac{1}{A^{n}} \rightarrow \frac{1}{(n-1) !} \int_{\tau_{U V}}^{\infty} d \tau \tau^{n-1} e^{-\tau A}.
\end{equation}
Following this, we do the following replacement:
\begin{equation}
\frac{1}{E}=\frac{1}{\sqrt{p^2+M^2}}\to \int_{\tau_{UV}}^{\infty} d\tau \frac{e^{-\tau(p^2+M^2)}}{\sqrt{\pi \tau}}. 
\end{equation}
Then the gap equation inside a rotating cylinder is obtained.

\begin{figure}[h]
\centering
\begin{minipage}{0.41\textwidth}
\centering
\includegraphics[width=1\linewidth]{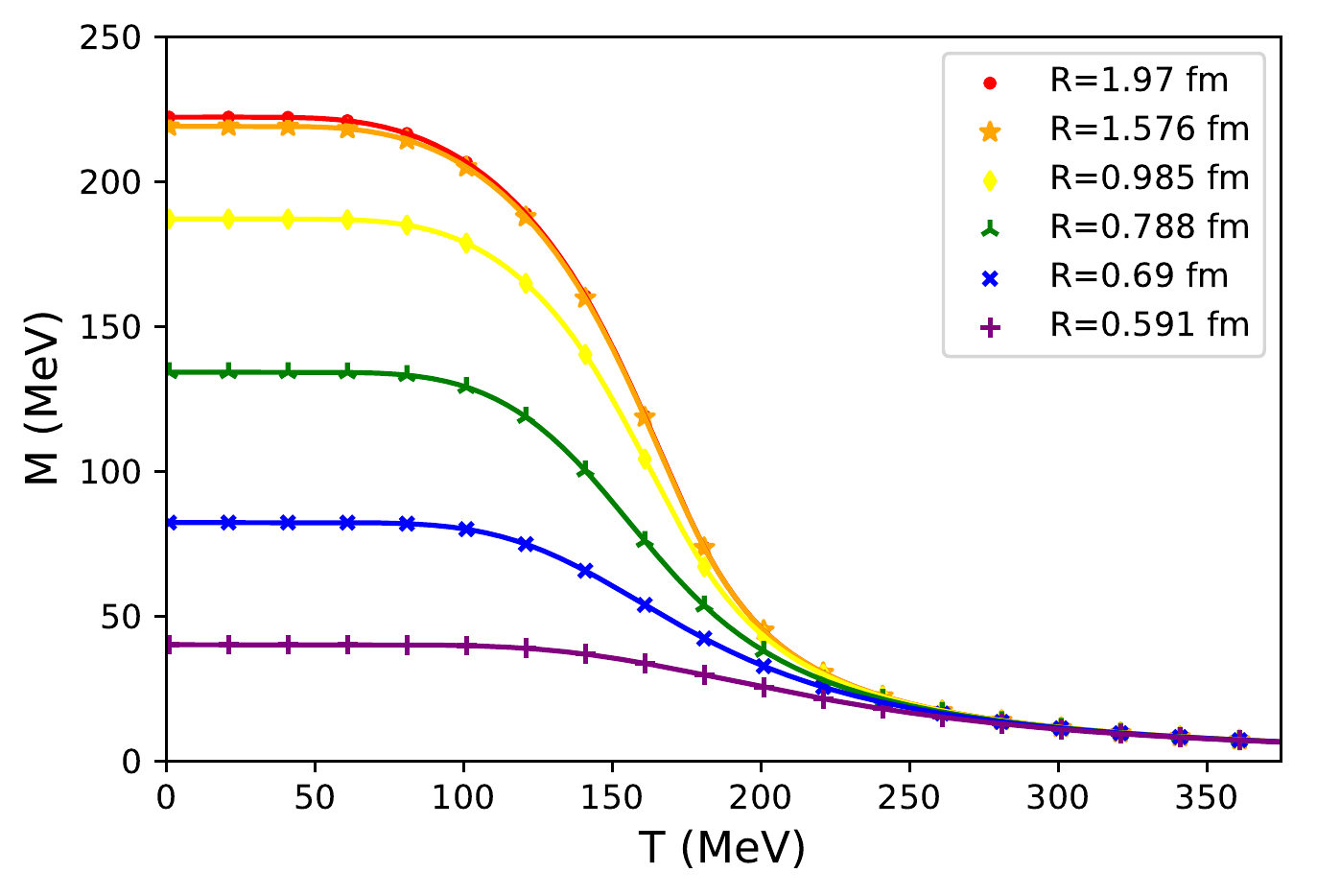} 
\end{minipage}
\caption{Effective mass $M$ as a function of $T$ in cylinders with different radius $R$, calculated at $r=0$ and $\Omega=0$.}
\label{fig1}
\end{figure}

In next several paragraphs, we investigate the chiral phase transition at a finite temperature and zero chemical potential. The parameters are taken as $m_0=5\ \mathrm{MeV}, G=3.26\times 10^{-6}\  \mathrm{MeV}^{-2}, \tau_{UV}=1/1080\  \mathrm{MeV}^{-2}$. First, we want to see how finite size influences the chiral phase transition at a finite temperature. Figure. \ref{fig1} presents the effective mass $M$ as a function of temperature $T$ inside cylinders with different radius $R$, calculated at $r=0$ with $\Omega=0$.  We can see that the finite size suppresses the effective mass and make the phase transition happens at a lower temperature. As we have mentioned, this is expected since spontaneous symmetry breaking only happens in infinite systems in principle, and thus a finite volume tends to partly restore the chiral symmetry. In Fig. \ref{fig1}, the curve of $R=1.97$ fm overlaps with the $M-T$ curve in infinite space, which means for the point $r=0$, $R=1.97 \  \mathrm{fm}$ is large enough to approach the infinite space when discussing the phase transition at a finite temperature.

\begin{figure}[h]
\centering
\begin{minipage}{0.41\textwidth}
\centering
\includegraphics[width=1\linewidth]{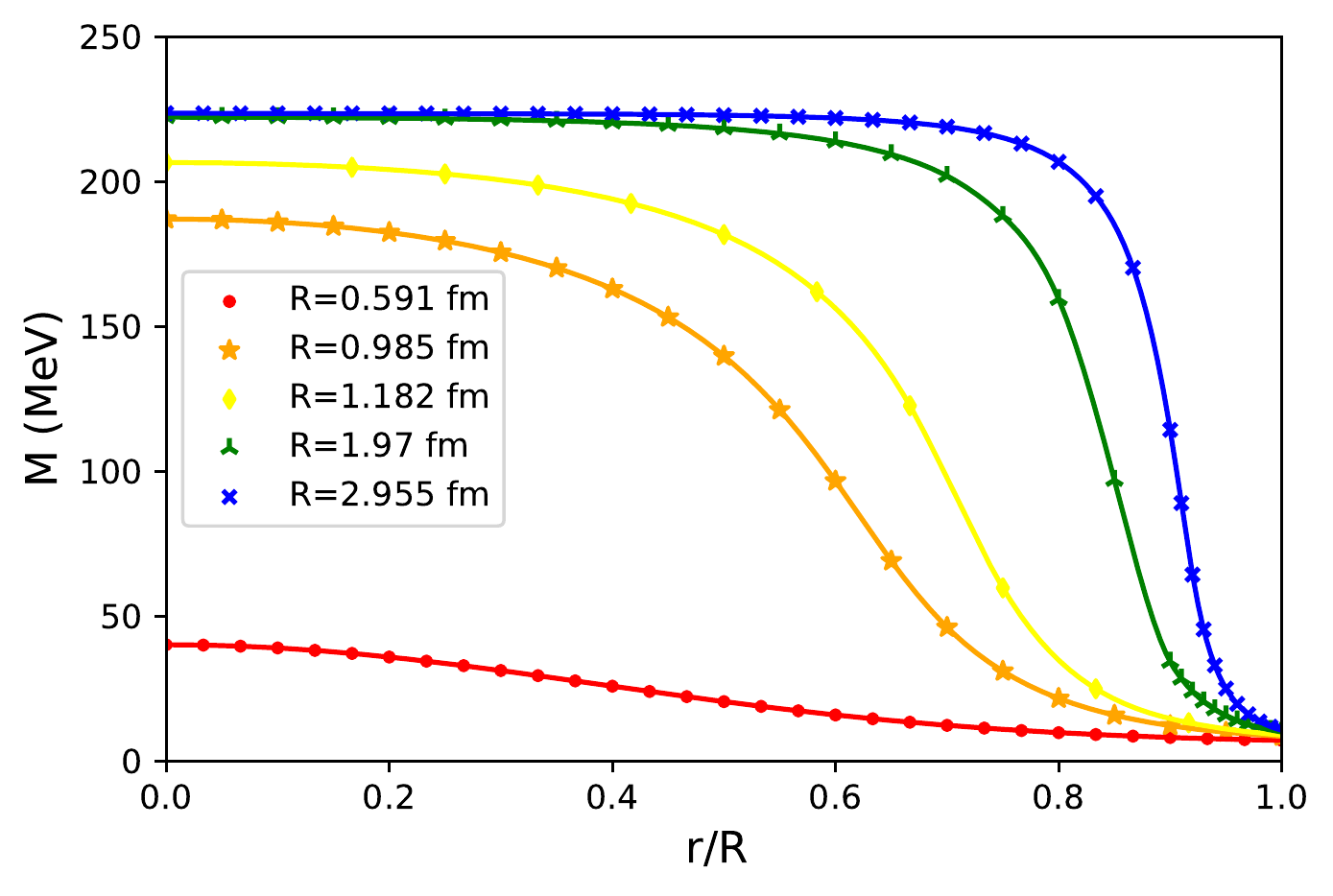} 
\end{minipage}
\caption{Effective mass $M$ as a function of $r/R$ in cylinders with different radius $R$, calculated at $T=0$ and $\Omega=0$.}
\label{fig2}
\end{figure}

To see the inhomogeneous effect induced by finite size, we plot Fig. \ref{fig2}, which presents the variation of effective mass with coordinate $r$ at $T=0$ and $\Omega=0$. We can see that the effective mass decreases with $r$. The condition which makes the local density approximation valid suggests that our results near the boundary (where the mass decreases sharply) are not very reliable, while the part far away from the boundary is reliable. An observation is that for large radius $R$, the effective mass remains almost a constant inside the cylinder, only decreases sharply near the boundary. Similar feature appeared in Ref. \cite{bound1}, although the regularization scheme in Ref. \cite{bound1} is different from ours. This feature is easy to understand because at large enough $R$, the boundary effects can be ignored unless we come very close to the boundary. 

\begin{figure}[h]
\centering
\begin{minipage}{0.41\textwidth}
\centering
\includegraphics[width=1\linewidth]{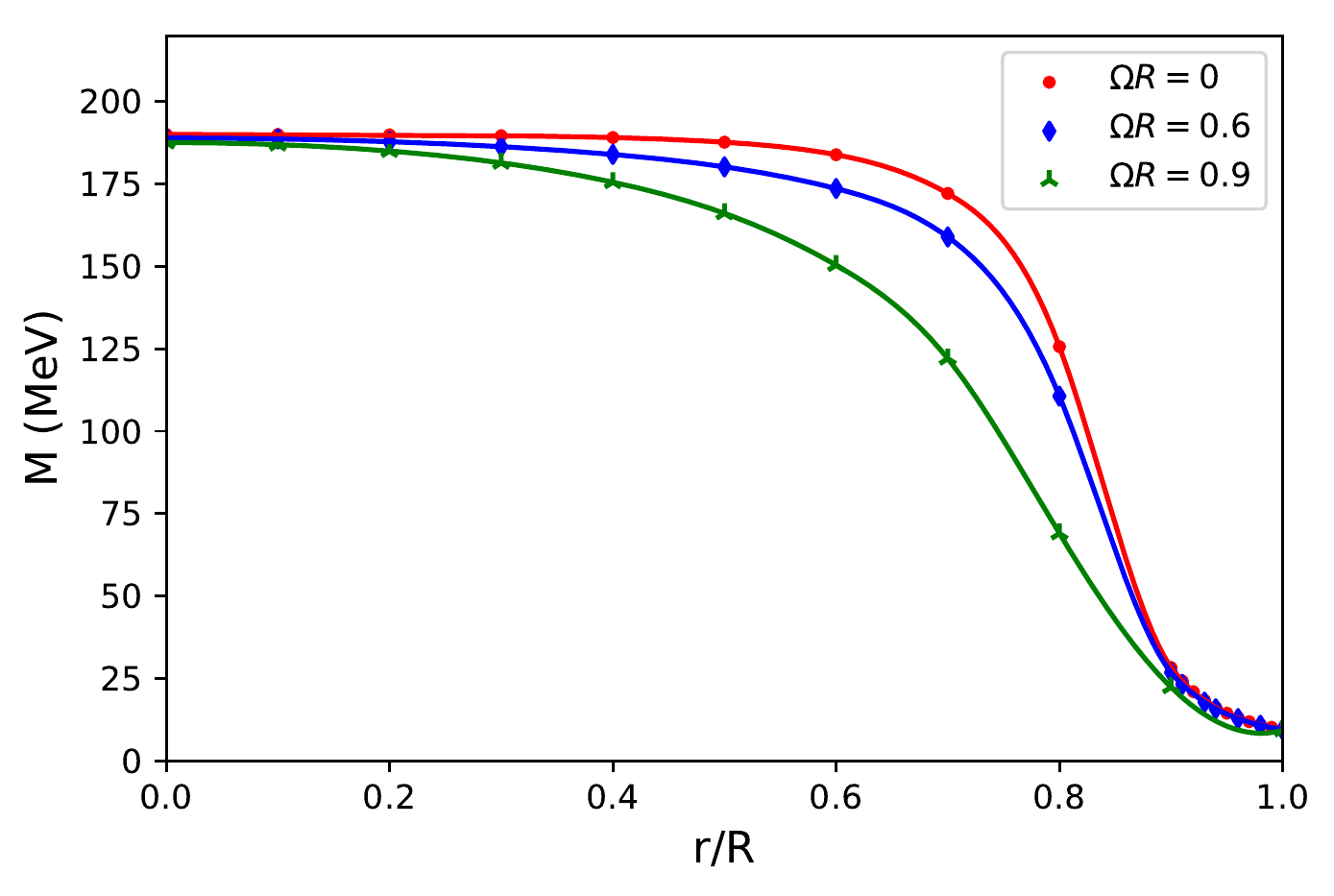} 
\end{minipage}
\caption{Effective mass $M$ as a function of $r/R$ under different angular velocity $\Omega$, calculated at $R=1.97\  \mathrm{fm}$ and $T=120\  \mathrm{MeV}$.}
\label{fig3}
\end{figure}

\begin{figure}[h]
\centering
\begin{minipage}{0.41\textwidth}
\centering
\includegraphics[width=1\linewidth]{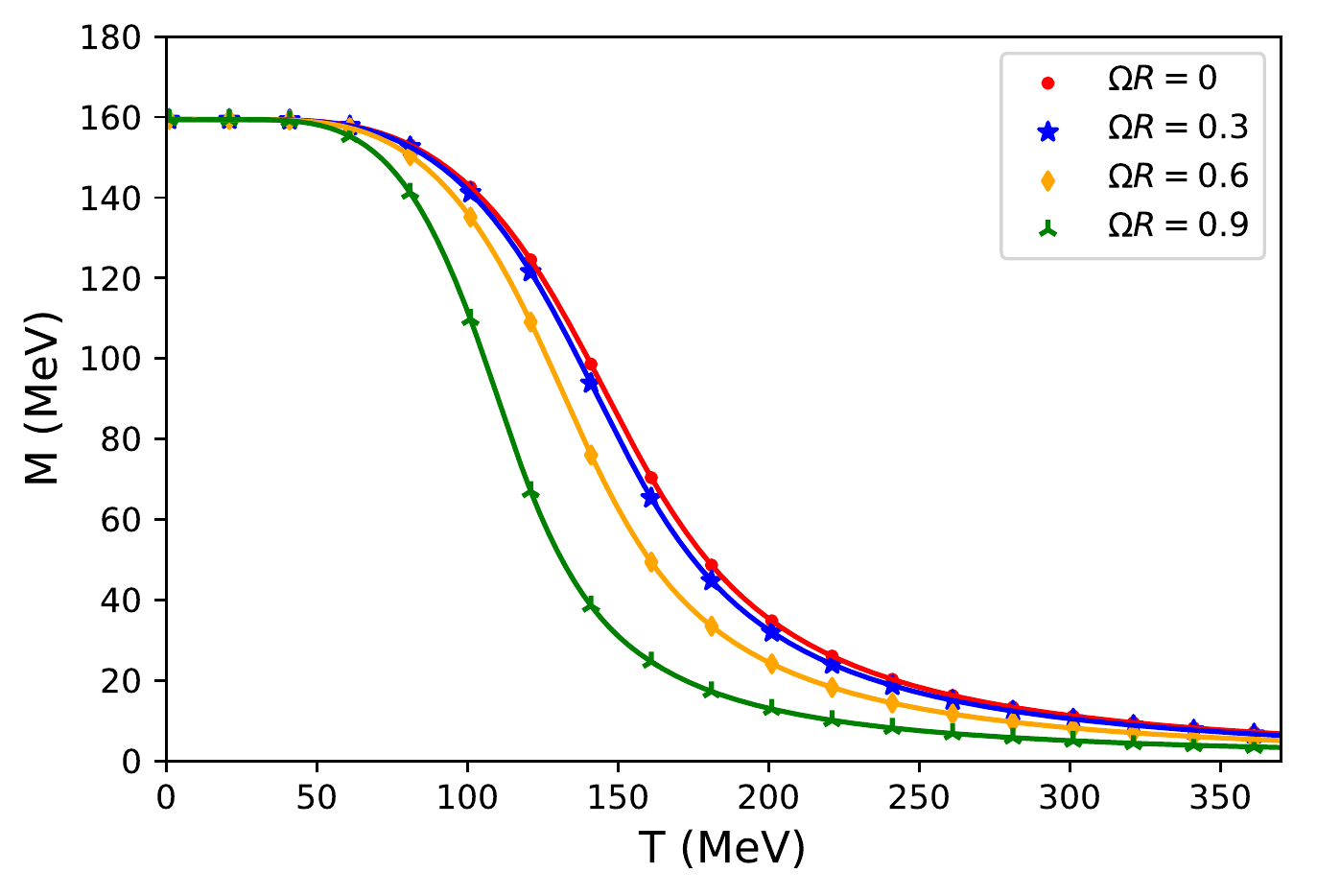} 
\end{minipage}
\caption{Effective mass $M$ as a function of $T$ under different angular velocity $\Omega$, calculated at $R=1.97\  \mathrm{fm}$ and $r=0.8 \ R$.}
\label{fig4}
\end{figure}

Now, let us investigate how rotation influences the effective mass. Figure. \ref{fig3} presents the variation of effective mass $M$ with $r$ under different angular velocity $\Omega$, calculated for $R=1.97\  \mathrm{fm}$ and $T=120 \ \mathrm{MeV}$ (Recall that rotation has no effect at zero temprature).
One can observe that rotation suppresses the effective mass. As we have said, rotation is similar with chemical potential, so this result is expected. Another observation is that the center point is less influenced by the rotation, while the "middle region" is more influenced. (We do not discuss the regions near the boundary since the results are not very reliable.) This may be understood intuitively that at the center the linear velocity $v=\Omega r$ is small and thus the effects caused by rotation is weak. In Figure. \ref{fig4}, we present the $M-T$ curve under different angular velocity $\Omega$, calculate for $R=1.97\  \mathrm{fm}$ and $r=0.8 \ R$. We can see that the rotation reduces the chiral transition temperature, which is also expected from the analogy between rotation and chemical potential. It is interesting to compare our results with that in Ref. \cite{YinJiang}, where the rotating cylinder is unbounded. The effects of rotation in our model are much smaller than that in Ref. \cite{YinJiang}, where the rotation can change the effective mass greatly and even restore the chiral symmetry. This difference comes from the fact that for an unbounded system, any angular velocity will cause faster than light. For example, in our model, $R=1.97$ fm is large enough to be seen as infinite volume. To avoid faster than light, the angular velocity cannot exceed $0.1$ GeV for this system. But the angular velocity adopted in Ref. \cite{YinJiang} reaches $0.8$ GeV, so it is understandable why the effects of rotation are so strong in Ref. \cite{YinJiang}. The discussions above tell us it is important to avoid the faster than light when investigating rotating systems. 

\begin{figure}[h]
\centering
\begin{minipage}{0.41\textwidth}
\centering
\includegraphics[width=1\linewidth]{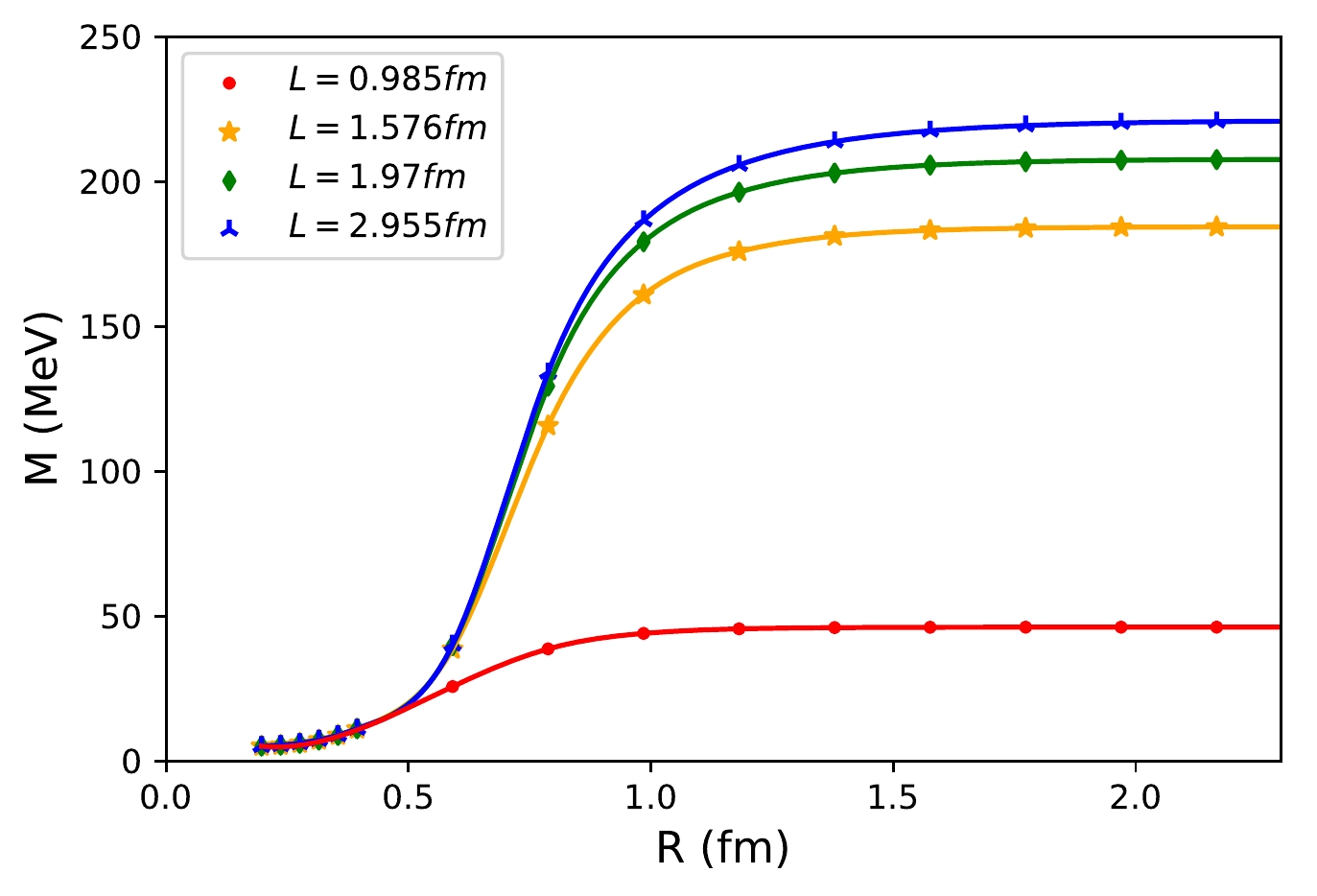} 
\end{minipage}
\caption{Effective mass as a function of radius $R$ for different length $L$, calculated at $r=0, T=0$ and $\Omega=0$.}
\label{fig5}
\end{figure}

\begin{figure}[h]
\centering
\begin{minipage}{0.41\textwidth}
\centering
\includegraphics[width=1\linewidth]{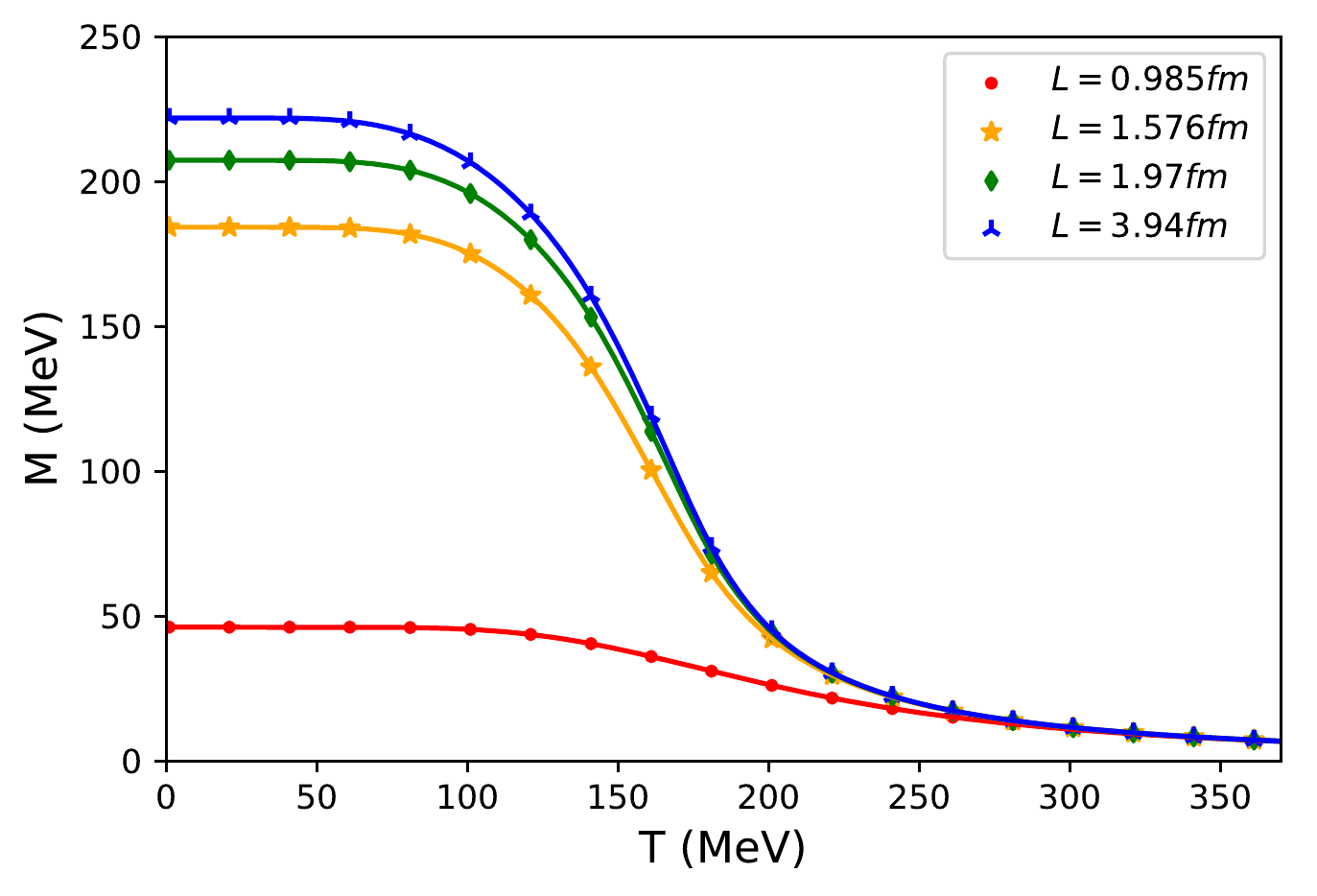} 
\end{minipage}
\caption{Effective mass as a function of $T$ for different length $L$, calculated at $R=1.97 \ \mathrm{fm}, T=0$ and $\Omega=0$.}
\label{fig6}
\end{figure}

The above calculations are done for infinitely long cylinder, but a finite long cylinder is more close to real systems. To bound the system at $z$ direction, we can impose the antiperiodic boundary condition at this direction, then the longitude momentum $k$ is discretized as
\begin{equation}
k=\frac{2\pi}{L}(n+\frac{1}{2}), \  \ n=0,\pm1,\pm2,...,
\end{equation}
where $L$ is the length of the cylinder. By replacing the integral over $k$ in Eq. (\ref{gap2}) by a discrete sum, we can investigate how the length influences the phase transition.
In Figure. \ref{fig5}, we present the effective mass as a function of radius $R$ for different length $L$, calculated at $r=0, T=0$ and $\Omega=0$.  First, one notices that the effective mass increases with $R$, which was found in Fig. \ref{fig1}. Second, one notice that the effective mass approaches a constant limit as $R$ increases, which indicates $R$ is large enough. But for different length $L$, this limit is different, which reflects the influence of $L$. In fact, for infinite $R$ and finite $L$, the system becomes a film. The thickness $L$ surely influences the effective mass, as shown by previous study \cite{qingwu}. Here we also would like to note that $L=3\  \mathrm{fm}$ is large enough that the outmost $M-R$ curve in Fig. \ref{fig5} is nearly overlap the curve with infinite $L$ (we do not plot). In Figure. \ref{fig6}, we present how the length $L$ influences the chiral phase transition. It can be seen that a finite $L$ has similar effect with a finite $R$, as one expected.

\begin{figure}[h]
\centering
\begin{minipage}{0.41\textwidth}
\centering
\includegraphics[width=1\linewidth]{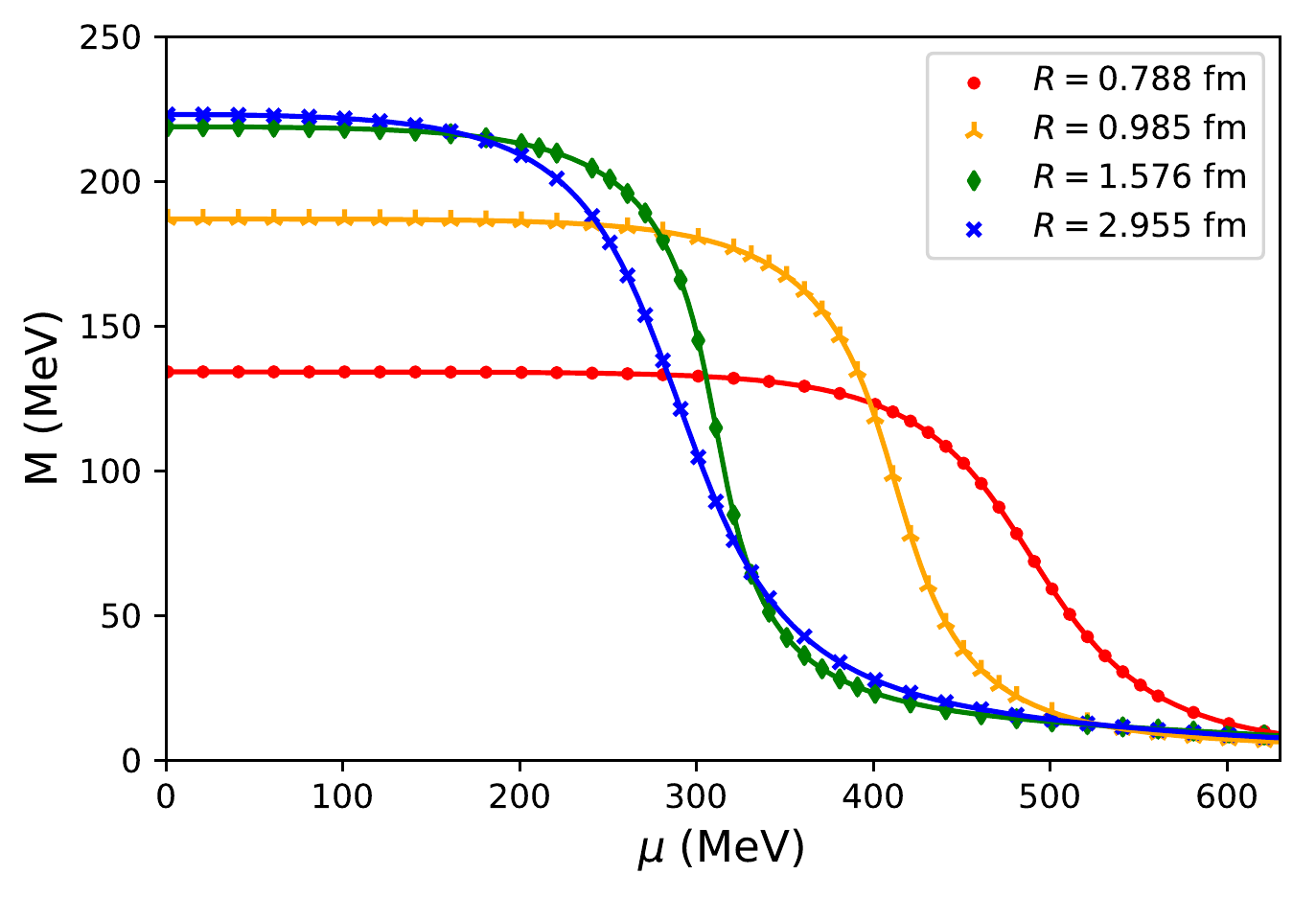} 
\end{minipage}
\caption{Effective mass as a function of chemical potential $\mu$ for different radius $R$, calculated at $r=0, T=50\  \mathrm{MeV}$ and $\Omega=0$.}
\label{fig7}
\end{figure}

\begin{figure}[h]
\centering
\begin{minipage}{0.41\textwidth}
\centering
\includegraphics[width=1\linewidth]{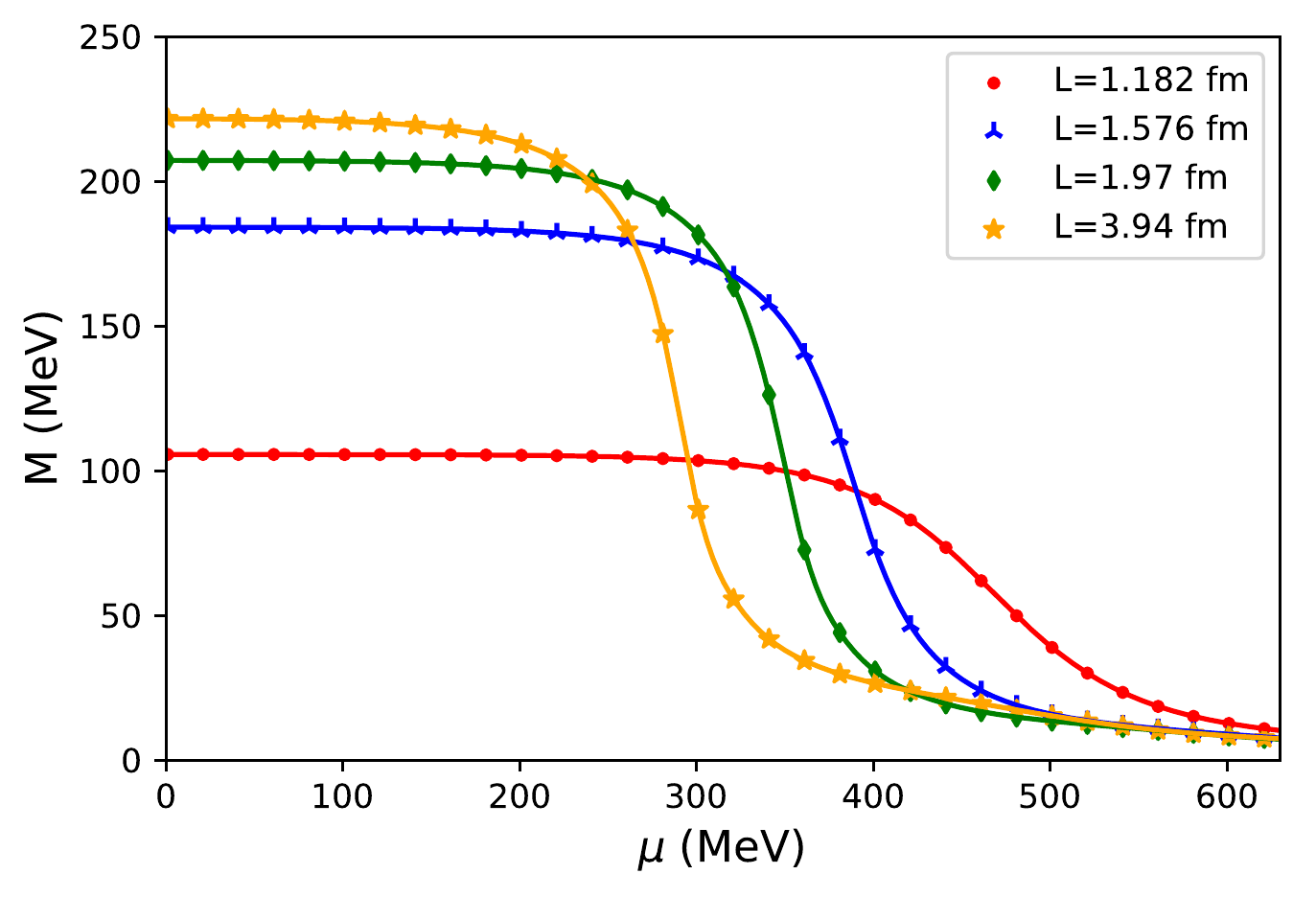} 
\end{minipage}
\caption{Effective mass as a function of chemical potential $\mu$ for different length $L$, calculated at $r=0, T=50\  \mathrm{MeV}$ and $\Omega=0$.}
\label{fig8}
\end{figure}

\begin{figure}[h]
\centering
\begin{minipage}{0.41\textwidth}
\centering
\includegraphics[width=1\linewidth]{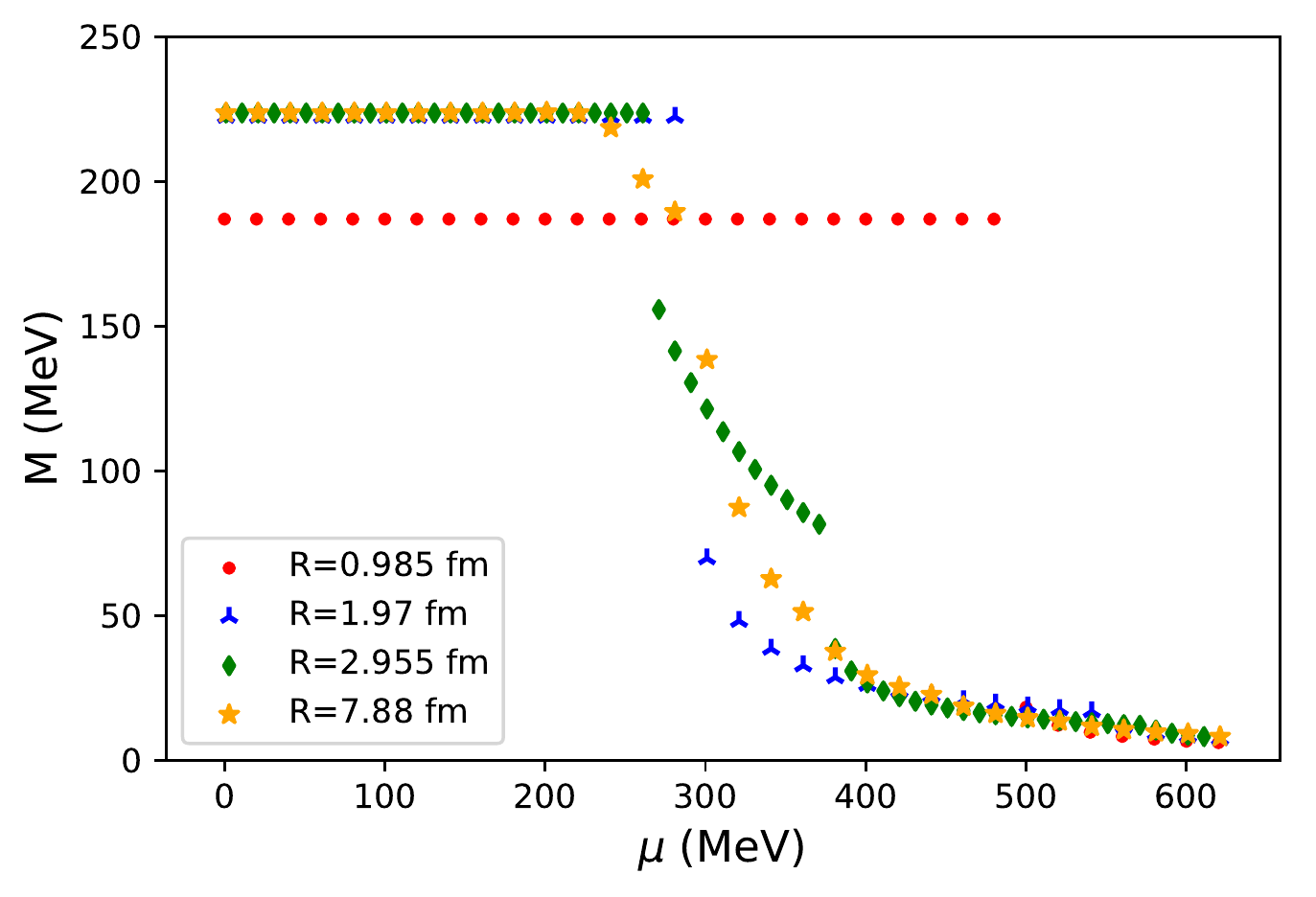} 
\end{minipage}
\caption{Effective mass as a function of chemical potential $\mu$ for different radius $R$, calculated at $r=0, T=0\  \mathrm{MeV}$ and $\Omega=0$.}
\label{fig9}
\end{figure}

Finally, let us discuss the chiral phase transition at a finite chemical potential.  Figures. \ref{fig7}-\ref{fig8} present how finite size influences the chiral phase transition at a finite chemical potential.  Contrast to the case of finite temperature, finite size will raise the chiral transition chemical potential, while at small chemical potential, finite size still suppresses the effective mass. We know that finite temperature will decrease the transition chemical potential. So in this aspect, small size does not equal to high temperature. Here, we especially want to present the case of zero temperature, in which we find some unusual behavior. In Fig. \ref{fig9}, we find at zero temperature, the phase transition at a finite chemical potential in a finite volume becomes discontinuous, while the infinite limit is a crossover in our regularization scheme. We also find there can be complex behavior of the $M-\mu$ curve at a "middle volume", see the curve of $R=2.955\  \mathrm{fm}$. Moreover, we find that, a size which is large enough for the phase transition at a finite temperature  may still be too small for the phase transition at a finite chemical potential at zero temperature (or more widely, at low temperature). For example, $R=1.97\  \mathrm{fm}$ is large enough for the phase transition at a finite temperature at $r=0$, but in Fig. \ref{fig9}, it is not large enough. These novel behaviors are partly related to the discretized momentum determined by the boundary condition. They could be unphysical due to the inapplicability of our model at finite chemical potential when temperature is very low, which certainly deserves futher investigation.

\begin{figure}[h]
\centering
\begin{minipage}{0.41\textwidth}
\centering
\includegraphics[width=1\linewidth]{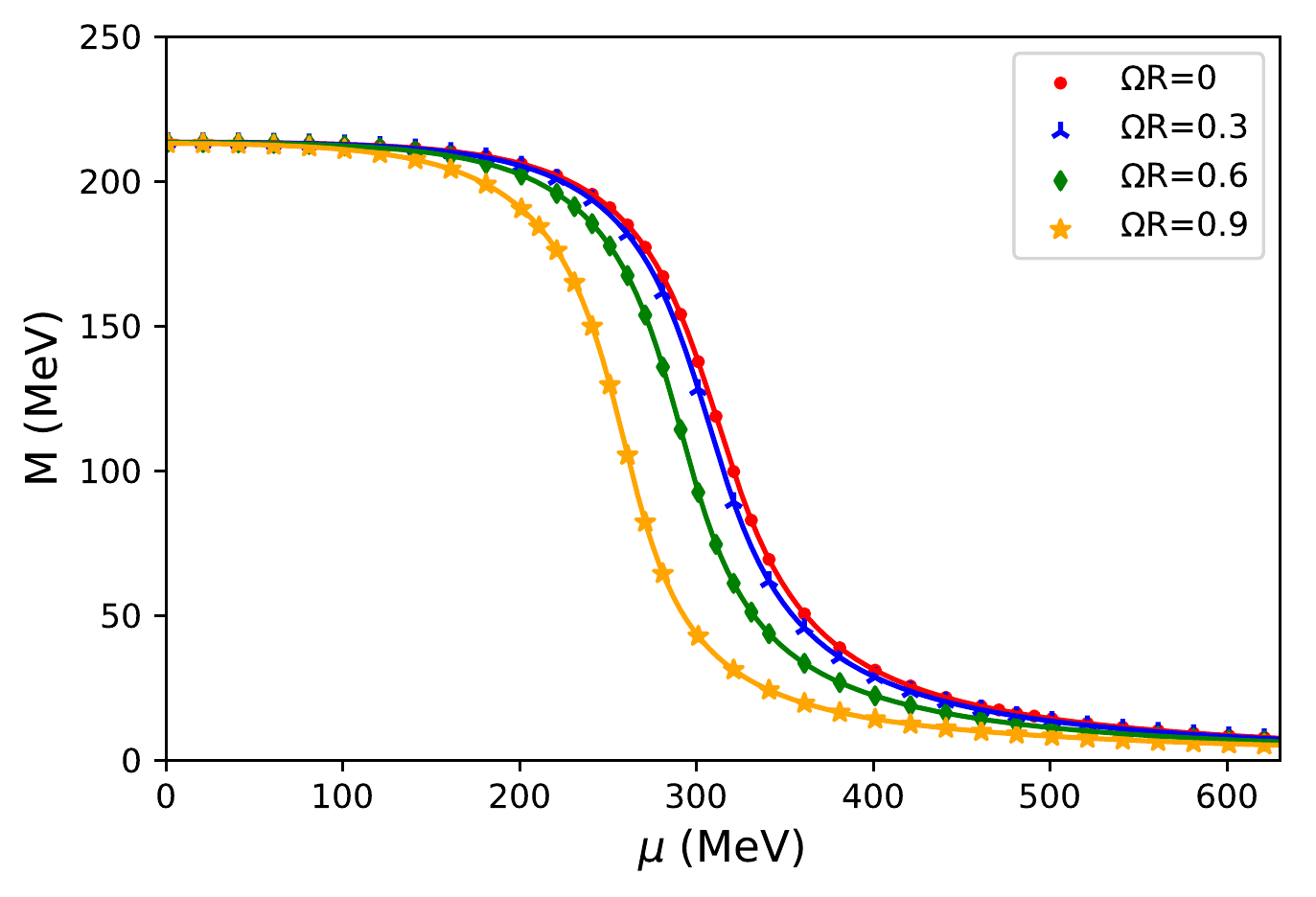} 
\end{minipage}
\caption{Effective mass as a function of chemical potential $\mu$ for different angular velocity $\Omega$, calculated for $R=1.97 \ \mathrm{fm}, r=0.6 \ R$ and $T=50 \ \mathrm{MeV}$.}
\label{fig10}
\end{figure}

 To show the influence of rotation on the phase transition at a finite chemical potential, we plot Figure. \ref{fig10}. We can observe that rotation reduces the chiral transition chemical potential, which is similar to the finite temperature case. This can also be understood by the similarity between rotation and chemical potential.

\section{Summary and discussion}\label{IV}

In this paper, we study the two-flavor NJL model in a rotating cylinder with spectral boundary condition. The effects of finite size and rotation on the chiral phase transition are investigated. It is found that finite size can lower the chiral transition temperature and raises the chiral transition chemical potential, while the rotation reduces both. By taking into account the inhomogeneous effects induced by finite size and rotation, we find that the effective mass decreases with coordinate $r$. It is worthwhile to discuss how our results are related to the heavy-ion collisions. In our model, when the temperature is not too low, the finite size effects are negligible when the size is larger than $3 \ \mathrm{fm}$, which is similar to the results obtained for box geometry with antiperiodic boundary conditions \cite{qingwu,DS4}.  But when the temperature is very low, finite size may have larger effects for the phase transition at finite chemical potential. It means that in heavy-ion collisions, where the matter size is about between 2-10 fm \cite{Palhares2011}, there may be much less finite size effects, since the QGP systems has high temperature and low density. For rotation, the influence is obvious only when $\Omega R$ exceeds 0.3, which is different from the results in Ref. \cite{FukushimaH}, where $\Omega R\sim 10^{-3}$ is large enough to have significant effects. The mostly rapidly rotating neutron star can reach $\Omega R=0.1$, so our calculation suggests it is reasonable to neglect the influence of rotation on the equation of state of neutron stars and treat it as a global effect \cite{rotatingstar}. But we should note, there still can be possibility that rotation has non-negligible effects on equation of state when considering its interplay with other factors such as the strong magnetic field. Finally, we emphasize that the boundary condition is important for finite size systems. The spectral boundary condition used here may not be realistic for systems such as the quark-gluon plasma created in heavy-ion collisions, and more careful investigations are needed.

\section*{Acknowledgements}
This work is supported in part by the National Natural Science Foundation of China (under Grants No. 12075117, No. 11535005, No. 11905104, No. 11690030 and No. 117751125) and by Nation Major State Basic Research and Development of China (2016YFE0129300). X. Luo is supported by the National Key Research and Development Program of China (2018YFE0205201),  the National Natural Science Foundation of China (Grants No. 11828501, No. 11890711 and No. 11861131009).

\bibliography{ref}

\begin{thebibliography}{27}%
\makeatletter
\providecommand \@ifxundefined [1]{%
 \@ifx{#1\undefined}
}%
\providecommand \@ifnum [1]{%
 \ifnum #1\expandafter \@firstoftwo
 \else \expandafter \@secondoftwo
 \fi
}%
\providecommand \@ifx [1]{%
 \ifx #1\expandafter \@firstoftwo
 \else \expandafter \@secondoftwo
 \fi
}%
\providecommand \natexlab [1]{#1}%
\providecommand \enquote  [1]{``#1''}%
\providecommand \bibnamefont  [1]{#1}%
\providecommand \bibfnamefont [1]{#1}%
\providecommand \citenamefont [1]{#1}%
\providecommand \href@noop [0]{\@secondoftwo}%
\providecommand \href [0]{\begingroup \@sanitize@url \@href}%
\providecommand \@href[1]{\@@startlink{#1}\@@href}%
\providecommand \@@href[1]{\endgroup#1\@@endlink}%
\providecommand \@sanitize@url [0]{\catcode `\\12\catcode `\$12\catcode
  `\&12\catcode `\#12\catcode `\^12\catcode `\_12\catcode `\%12\relax}%
\providecommand \@@startlink[1]{}%
\providecommand \@@endlink[0]{}%
\providecommand \url  [0]{\begingroup\@sanitize@url \@url }%
\providecommand \@url [1]{\endgroup\@href {#1}{\urlprefix }}%
\providecommand \urlprefix  [0]{URL }%
\providecommand \Eprint [0]{\href }%
\providecommand \doibase [0]{http://dx.doi.org/}%
\providecommand \selectlanguage [0]{\@gobble}%
\providecommand \bibinfo  [0]{\@secondoftwo}%
\providecommand \bibfield  [0]{\@secondoftwo}%
\providecommand \translation [1]{[#1]}%
\providecommand \BibitemOpen [0]{}%
\providecommand \bibitemStop [0]{}%
\providecommand \bibitemNoStop [0]{.\EOS\space}%
\providecommand \EOS [0]{\spacefactor3000\relax}%
\providecommand \BibitemShut  [1]{\csname bibitem#1\endcsname}%
\let\auto@bib@innerbib\@empty
\bibitem [{\citenamefont {Adamczyk}\ \emph {et~al.}(2017)\citenamefont
  {Adamczyk}, \citenamefont {Adkins}, \citenamefont {Agakishiev} \emph
  {et~al.}}]{vorticalfluid}%
  \BibitemOpen
  \bibfield  {author} {\bibinfo {author} {\bibfnamefont {L.}~\bibnamefont
  {Adamczyk}}, \bibinfo {author} {\bibfnamefont {J.~K.}\ \bibnamefont
  {Adkins}}, \bibinfo {author} {\bibfnamefont {G.}~\bibnamefont {Agakishiev}},
  \emph {et~al.},\ }\href {\doibase 10.1038/nature23004} {\bibfield  {journal}
  {\bibinfo  {journal} {Nature(London)}\ }\textbf {\bibinfo {volume} {548}},\
  \bibinfo {pages} {62} (\bibinfo {year} {2017})}\BibitemShut {NoStop}%
\bibitem [{\citenamefont {Watts}\ \emph {et~al.}(2016)\citenamefont {Watts},
  \citenamefont {Andersson} \emph {et~al.}}]{starreview}%
  \BibitemOpen
  \bibfield  {author} {\bibinfo {author} {\bibfnamefont {A.~L.}\ \bibnamefont
  {Watts}}, \bibinfo {author} {\bibfnamefont {N.}~\bibnamefont {Andersson}},
  \emph {et~al.},\ }\href {\doibase 10.1103/RevModPhys.88.021001} {\bibfield
  {journal} {\bibinfo  {journal} {Rev. Mod. Phys.}\ }\textbf {\bibinfo {volume}
  {88}},\ \bibinfo {pages} {021001} (\bibinfo {year} {2016})}\BibitemShut
  {NoStop}%
\bibitem [{\citenamefont {Kharzeev}\ and\ \citenamefont
  {Zhitnitsky}(2007)}]{chiralv1}%
  \BibitemOpen
  \bibfield  {author} {\bibinfo {author} {\bibfnamefont {D.}~\bibnamefont
  {Kharzeev}}\ and\ \bibinfo {author} {\bibfnamefont {A.}~\bibnamefont
  {Zhitnitsky}},\ }\href {\doibase 10.1016/j.nuclphysa.2007.10.001} {\bibfield
  {journal} {\bibinfo  {journal} {Nucl. Phys. A}\ }\textbf {\bibinfo {volume}
  {797}},\ \bibinfo {pages} {67} (\bibinfo {year} {2007})}\BibitemShut
  {NoStop}%
\bibitem [{\citenamefont {Son}\ and\ \citenamefont
  {Sur\'owka}(2009)}]{chiralv2}%
  \BibitemOpen
  \bibfield  {author} {\bibinfo {author} {\bibfnamefont {D.~T.}\ \bibnamefont
  {Son}}\ and\ \bibinfo {author} {\bibfnamefont {P.}~\bibnamefont
  {Sur\'owka}},\ }\href {\doibase 10.1103/PhysRevLett.103.191601} {\bibfield
  {journal} {\bibinfo  {journal} {Phys. Rev. Lett.}\ }\textbf {\bibinfo
  {volume} {103}},\ \bibinfo {pages} {191601} (\bibinfo {year}
  {2009})}\BibitemShut {NoStop}%
\bibitem [{\citenamefont {Kharzeev}\ and\ \citenamefont
  {Son}(2011)}]{chiralv3}%
  \BibitemOpen
  \bibfield  {author} {\bibinfo {author} {\bibfnamefont {D.~E.}\ \bibnamefont
  {Kharzeev}}\ and\ \bibinfo {author} {\bibfnamefont {D.~T.}\ \bibnamefont
  {Son}},\ }\href {\doibase 10.1103/PhysRevLett.106.062301} {\bibfield
  {journal} {\bibinfo  {journal} {Phys. Rev. Lett.}\ }\textbf {\bibinfo
  {volume} {106}},\ \bibinfo {pages} {062301} (\bibinfo {year}
  {2011})}\BibitemShut {NoStop}%
\bibitem [{\citenamefont {Stephanov}\ and\ \citenamefont {Yin}(2012)}]{Yinke}%
  \BibitemOpen
  \bibfield  {author} {\bibinfo {author} {\bibfnamefont {M.~A.}\ \bibnamefont
  {Stephanov}}\ and\ \bibinfo {author} {\bibfnamefont {Y.}~\bibnamefont
  {Yin}},\ }\href {\doibase 10.1103/PhysRevLett.109.162001} {\bibfield
  {journal} {\bibinfo  {journal} {Phys. Rev. Lett.}\ }\textbf {\bibinfo
  {volume} {109}},\ \bibinfo {pages} {162001} (\bibinfo {year}
  {2012})}\BibitemShut {NoStop}%
\bibitem [{\citenamefont {Jiang}\ and\ \citenamefont {Liao}(2016)}]{YinJiang}%
  \BibitemOpen
  \bibfield  {author} {\bibinfo {author} {\bibfnamefont {Y.}~\bibnamefont
  {Jiang}}\ and\ \bibinfo {author} {\bibfnamefont {J.}~\bibnamefont {Liao}},\
  }\href {\doibase 10.1103/PhysRevLett.117.192302} {\bibfield  {journal}
  {\bibinfo  {journal} {Phys. Rev. Lett.}\ }\textbf {\bibinfo {volume} {117}},\
  \bibinfo {pages} {192302} (\bibinfo {year} {2016})}\BibitemShut {NoStop}%
\bibitem [{\citenamefont {Ebihara}\ \emph {et~al.}(2017)\citenamefont
  {Ebihara}, \citenamefont {Fukushima},\ and\ \citenamefont {Mameda}}]{bound1}%
  \BibitemOpen
  \bibfield  {author} {\bibinfo {author} {\bibfnamefont {S.}~\bibnamefont
  {Ebihara}}, \bibinfo {author} {\bibfnamefont {K.}~\bibnamefont {Fukushima}},
  \ and\ \bibinfo {author} {\bibfnamefont {K.}~\bibnamefont {Mameda}},\ }\href
  {\doibase https://doi.org/10.1016/j.physletb.2016.11.010} {\bibfield
  {journal} {\bibinfo  {journal} {Phys. Lett. B}\ }\textbf {\bibinfo {volume}
  {764}},\ \bibinfo {pages} {94 } (\bibinfo {year} {2017})}\BibitemShut
  {NoStop}%
\bibitem [{\citenamefont {Chernodub}\ and\ \citenamefont
  {Gongyo}()}]{cylinder}%
  \BibitemOpen
  \bibfield  {author} {\bibinfo {author} {\bibfnamefont {M.~N.}\ \bibnamefont
  {Chernodub}}\ and\ \bibinfo {author} {\bibfnamefont {S.}~\bibnamefont
  {Gongyo}},\ }\href {\doibase 10.1007/JHEP01(2017)136} {\bibfield  {journal}
  {\bibinfo  {journal} {J. High Energy Phys.}\ }\textbf {\bibinfo {volume}
  {01}},\ \bibinfo {pages} {(2017) 136}}\BibitemShut {NoStop}%
\bibitem [{\citenamefont {Chen}\ \emph {et~al.}(2016)\citenamefont {Chen},
  \citenamefont {Fukushima}, \citenamefont {Huang},\ and\ \citenamefont
  {Mameda}}]{FukushimaH}%
  \BibitemOpen
  \bibfield  {author} {\bibinfo {author} {\bibfnamefont {H.-L.}\ \bibnamefont
  {Chen}}, \bibinfo {author} {\bibfnamefont {K.}~\bibnamefont {Fukushima}},
  \bibinfo {author} {\bibfnamefont {X.-G.}\ \bibnamefont {Huang}}, \ and\
  \bibinfo {author} {\bibfnamefont {K.}~\bibnamefont {Mameda}},\ }\href
  {\doibase 10.1103/PhysRevD.93.104052} {\bibfield  {journal} {\bibinfo
  {journal} {Phys. Rev. D}\ }\textbf {\bibinfo {volume} {93}},\ \bibinfo
  {pages} {104052} (\bibinfo {year} {2016})}\BibitemShut {NoStop}%
\bibitem [{\citenamefont {McInnes}(2016)}]{holo3}%
  \BibitemOpen
  \bibfield  {author} {\bibinfo {author} {\bibfnamefont {B.}~\bibnamefont
  {McInnes}},\ }\href {\doibase
  https://doi.org/10.1016/j.nuclphysb.2016.08.001} {\bibfield  {journal}
  {\bibinfo  {journal} {Nucl. Phys. B}\ }\textbf {\bibinfo {volume} {911}},\
  \bibinfo {pages} {173 } (\bibinfo {year} {2016})}\BibitemShut {NoStop}%
\bibitem [{\citenamefont {Klein}(2017)}]{reviewK}%
  \BibitemOpen
  \bibfield  {author} {\bibinfo {author} {\bibfnamefont {B.}~\bibnamefont
  {Klein}},\ }\href {\doibase https://doi.org/10.1016/j.physrep.2017.09.002}
  {\bibfield  {journal} {\bibinfo  {journal} {Phys. Rep.}\ }\textbf {\bibinfo
  {volume} {707-708}},\ \bibinfo {pages} {1 } (\bibinfo {year}
  {2017})}\BibitemShut {NoStop}%
\bibitem [{\citenamefont {Zhang}\ \emph
  {et~al.}(2020{\natexlab{a}})\citenamefont {Zhang}, \citenamefont {Shi},\ and\
  \citenamefont {Zong}}]{zzhang1}%
  \BibitemOpen
  \bibfield  {author} {\bibinfo {author} {\bibfnamefont {Z.}~\bibnamefont
  {Zhang}}, \bibinfo {author} {\bibfnamefont {C.}~\bibnamefont {Shi}}, \ and\
  \bibinfo {author} {\bibfnamefont {H.-S.}\ \bibnamefont {Zong}},\ }\href
  {\doibase 10.1103/PhysRevD.101.043006} {\bibfield  {journal} {\bibinfo
  {journal} {Phys. Rev. D}\ }\textbf {\bibinfo {volume} {101}},\ \bibinfo
  {pages} {043006} (\bibinfo {year} {2020}{\natexlab{a}})}\BibitemShut
  {NoStop}%
\bibitem [{\citenamefont {Zhang}\ \emph
  {et~al.}(2020{\natexlab{b}})\citenamefont {Zhang}, \citenamefont {Shi},
  \citenamefont {Luo},\ and\ \citenamefont {Zong}}]{zzhang2}%
  \BibitemOpen
  \bibfield  {author} {\bibinfo {author} {\bibfnamefont {Z.}~\bibnamefont
  {Zhang}}, \bibinfo {author} {\bibfnamefont {C.}~\bibnamefont {Shi}}, \bibinfo
  {author} {\bibfnamefont {X.}~\bibnamefont {Luo}}, \ and\ \bibinfo {author}
  {\bibfnamefont {H.-S.}\ \bibnamefont {Zong}},\ }\href {\doibase
  10.1103/PhysRevD.101.074036} {\bibfield  {journal} {\bibinfo  {journal}
  {Phys. Rev. D}\ }\textbf {\bibinfo {volume} {101}},\ \bibinfo {pages}
  {074036} (\bibinfo {year} {2020}{\natexlab{b}})}\BibitemShut {NoStop}%
\bibitem [{\citenamefont {Chodos}\ \emph {et~al.}(1974)\citenamefont {Chodos},
  \citenamefont {Jaffe}, \citenamefont {Johnson}, \citenamefont {Thorn},\ and\
  \citenamefont {Weisskopf}}]{MIT1}%
  \BibitemOpen
  \bibfield  {author} {\bibinfo {author} {\bibfnamefont {A.}~\bibnamefont
  {Chodos}}, \bibinfo {author} {\bibfnamefont {R.~L.}\ \bibnamefont {Jaffe}},
  \bibinfo {author} {\bibfnamefont {K.}~\bibnamefont {Johnson}}, \bibinfo
  {author} {\bibfnamefont {C.~B.}\ \bibnamefont {Thorn}}, \ and\ \bibinfo
  {author} {\bibfnamefont {V.~F.}\ \bibnamefont {Weisskopf}},\ }\href {\doibase
  10.1103/PhysRevD.9.3471} {\bibfield  {journal} {\bibinfo  {journal} {Phys.
  Rev. D}\ }\textbf {\bibinfo {volume} {9}},\ \bibinfo {pages} {3471} (\bibinfo
  {year} {1974})}\BibitemShut {NoStop}%
\bibitem [{\citenamefont {Horta\c{c}su}\ \emph {et~al.}(1980)\citenamefont
  {Horta\c{c}su}, \citenamefont {Rothe},\ and\ \citenamefont
  {Schroer}}]{spectral}%
  \BibitemOpen
  \bibfield  {author} {\bibinfo {author} {\bibfnamefont {M.}~\bibnamefont
  {Horta\c{c}su}}, \bibinfo {author} {\bibfnamefont {K.}~\bibnamefont {Rothe}},
  \ and\ \bibinfo {author} {\bibfnamefont {B.}~\bibnamefont {Schroer}},\ }\href
  {\doibase https://doi.org/10.1016/0550-3213(80)90384-3} {\bibfield  {journal}
  {\bibinfo  {journal} {Nucl. Phys. B}\ }\textbf {\bibinfo {volume} {171}},\
  \bibinfo {pages} {530 } (\bibinfo {year} {1980})}\BibitemShut {NoStop}%
\bibitem [{\citenamefont {Ambru\c{s}}\ and\ \citenamefont
  {Winstanley}(2016)}]{rcylinder}%
  \BibitemOpen
  \bibfield  {author} {\bibinfo {author} {\bibfnamefont {V.~E.}\ \bibnamefont
  {Ambru\c{s}}}\ and\ \bibinfo {author} {\bibfnamefont {E.}~\bibnamefont
  {Winstanley}},\ }\href {\doibase 10.1103/PhysRevD.93.104014} {\bibfield
  {journal} {\bibinfo  {journal} {Phys. Rev. D}\ }\textbf {\bibinfo {volume}
  {93}},\ \bibinfo {pages} {104014} (\bibinfo {year} {2016})}\BibitemShut
  {NoStop}%
\bibitem [{\citenamefont {Zhang}\ \emph
  {et~al.}(2020{\natexlab{c}})\citenamefont {Zhang}, \citenamefont {Shi},
  \citenamefont {Luo},\ and\ \citenamefont {Zong}}]{zzhang3}%
  \BibitemOpen
  \bibfield  {author} {\bibinfo {author} {\bibfnamefont {Z.}~\bibnamefont
  {Zhang}}, \bibinfo {author} {\bibfnamefont {C.}~\bibnamefont {Shi}}, \bibinfo
  {author} {\bibfnamefont {X.}~\bibnamefont {Luo}}, \ and\ \bibinfo {author}
  {\bibfnamefont {H.-S.}\ \bibnamefont {Zong}},\ }\href {\doibase
  10.1103/PhysRevD.102.065002} {\bibfield  {journal} {\bibinfo  {journal}
  {Phys. Rev. D}\ }\textbf {\bibinfo {volume} {102}},\ \bibinfo {pages}
  {065002} (\bibinfo {year} {2020}{\natexlab{c}})}\BibitemShut {NoStop}%
\bibitem [{\citenamefont {Wang}\ \emph {et~al.}(2018)\citenamefont {Wang},
  \citenamefont {Xia},\ and\ \citenamefont {Zong}}]{qingwu}%
  \BibitemOpen
  \bibfield  {author} {\bibinfo {author} {\bibfnamefont {Q.-W.}\ \bibnamefont
  {Wang}}, \bibinfo {author} {\bibfnamefont {Y.-H.}\ \bibnamefont {Xia}}, \
  and\ \bibinfo {author} {\bibfnamefont {H.-S.}\ \bibnamefont {Zong}},\ }\href
  {\doibase 10.1142/S0217732318502322} {\bibfield  {journal} {\bibinfo
  {journal} {Mod. Phys. Lett. A}\ }\textbf {\bibinfo {volume} {33}},\ \bibinfo
  {pages} {1850232} (\bibinfo {year} {2018})}\BibitemShut {NoStop}%
\bibitem [{\citenamefont {Xia}\ \emph {et~al.}(2019)\citenamefont {Xia},
  \citenamefont {Wang}, \citenamefont {Feng},\ and\ \citenamefont
  {Zong}}]{Xia}%
  \BibitemOpen
  \bibfield  {author} {\bibinfo {author} {\bibfnamefont {Y.-H.}\ \bibnamefont
  {Xia}}, \bibinfo {author} {\bibfnamefont {Q.-W.}\ \bibnamefont {Wang}},
  \bibinfo {author} {\bibfnamefont {H.-T.}\ \bibnamefont {Feng}}, \ and\
  \bibinfo {author} {\bibfnamefont {H.-S.}\ \bibnamefont {Zong}},\ }\href
  {\doibase 10.1088/1674-1137/43/3/034101} {\bibfield  {journal} {\bibinfo
  {journal} {Chin. Phys. C}\ }\textbf {\bibinfo {volume} {43}},\ \bibinfo
  {pages} {034101} (\bibinfo {year} {2019})}\BibitemShut {NoStop}%
\bibitem [{\citenamefont {Weinberg}(2013)}]{Weinberg2}%
  \BibitemOpen
  \bibfield  {author} {\bibinfo {author} {\bibfnamefont {S.}~\bibnamefont
  {Weinberg}},\ }\href@noop {} {\emph {\bibinfo {title} {{The Quantum Theory of
  Fields. Vol. 2: Modern Applications}}}}\ (\bibinfo  {publisher} {Cambridge
  University Press, Cambridge, England},\ \bibinfo {year} {2013})\BibitemShut
  {NoStop}%
\bibitem [{\citenamefont {Palhares}\ \emph {et~al.}(2011)\citenamefont
  {Palhares}, \citenamefont {Fraga},\ and\ \citenamefont
  {Kodama}}]{Palhares2011}%
  \BibitemOpen
  \bibfield  {author} {\bibinfo {author} {\bibfnamefont {L.~F.}\ \bibnamefont
  {Palhares}}, \bibinfo {author} {\bibfnamefont {E.~S.}\ \bibnamefont {Fraga}},
  \ and\ \bibinfo {author} {\bibfnamefont {T.}~\bibnamefont {Kodama}},\ }\href
  {\doibase 10.1088/0954-3899/38/8/085101} {\bibfield  {journal} {\bibinfo
  {journal} {J. Phys. G: Nucl. Part. Phys.}\ }\textbf {\bibinfo {volume}
  {38}},\ \bibinfo {pages} {085101} (\bibinfo {year} {2011})}\BibitemShut
  {NoStop}%
\bibitem [{\citenamefont {Zhao}\ \emph {et~al.}(2019)\citenamefont {Zhao},
  \citenamefont {Zhang}, \citenamefont {Zhang},\ and\ \citenamefont
  {Zong}}]{MRE4}%
  \BibitemOpen
  \bibfield  {author} {\bibinfo {author} {\bibfnamefont {Y.-P.}\ \bibnamefont
  {Zhao}}, \bibinfo {author} {\bibfnamefont {R.-R.}\ \bibnamefont {Zhang}},
  \bibinfo {author} {\bibfnamefont {H.}~\bibnamefont {Zhang}}, \ and\ \bibinfo
  {author} {\bibfnamefont {H.-S.}\ \bibnamefont {Zong}},\ }\href {\doibase
  10.1088/1674-1137/43/6/063101} {\bibfield  {journal} {\bibinfo  {journal}
  {Chin. Phys. C}\ }\textbf {\bibinfo {volume} {43}},\ \bibinfo {pages}
  {063101} (\bibinfo {year} {2019})}\BibitemShut {NoStop}%
\bibitem [{\citenamefont {Li}\ \emph {et~al.}(2019)\citenamefont {Li},
  \citenamefont {Cui}, \citenamefont {Zhou}, \citenamefont {An}, \citenamefont
  {Zhang},\ and\ \citenamefont {Zong}}]{DS2}%
  \BibitemOpen
  \bibfield  {author} {\bibinfo {author} {\bibfnamefont {B.-L.}\ \bibnamefont
  {Li}}, \bibinfo {author} {\bibfnamefont {Z.-F.}\ \bibnamefont {Cui}},
  \bibinfo {author} {\bibfnamefont {B.-W.}\ \bibnamefont {Zhou}}, \bibinfo
  {author} {\bibfnamefont {S.}~\bibnamefont {An}}, \bibinfo {author}
  {\bibfnamefont {L.-P.}\ \bibnamefont {Zhang}}, \ and\ \bibinfo {author}
  {\bibfnamefont {H.-S.}\ \bibnamefont {Zong}},\ }\href {\doibase
  https://doi.org/10.1016/j.nuclphysb.2018.11.015} {\bibfield  {journal}
  {\bibinfo  {journal} {Nucl. Phys. B}\ }\textbf {\bibinfo {volume} {938}},\
  \bibinfo {pages} {298 } (\bibinfo {year} {2019})}\BibitemShut {NoStop}%
\bibitem [{\citenamefont {Shi}\ \emph {et~al.}(2018)\citenamefont {Shi},
  \citenamefont {Xia}, \citenamefont {Jia},\ and\ \citenamefont {Zong}}]{DS4}%
  \BibitemOpen
  \bibfield  {author} {\bibinfo {author} {\bibfnamefont {C.}~\bibnamefont
  {Shi}}, \bibinfo {author} {\bibfnamefont {Y.-H.}\ \bibnamefont {Xia}},
  \bibinfo {author} {\bibfnamefont {W.-B.}\ \bibnamefont {Jia}}, \ and\
  \bibinfo {author} {\bibfnamefont {H.-S.}\ \bibnamefont {Zong}},\ }\href
  {\doibase 10.1007/s11433-017-9177-4} {\bibfield  {journal} {\bibinfo
  {journal} {Sci. China Phys, Mech.}\ }\textbf {\bibinfo {volume} {61}},\
  \bibinfo {pages} {082021} (\bibinfo {year} {2018})}\BibitemShut {NoStop}%
\bibitem [{\citenamefont {Klevansky}(1992)}]{NJLreview}%
  \BibitemOpen
  \bibfield  {author} {\bibinfo {author} {\bibfnamefont {S.~P.}\ \bibnamefont
  {Klevansky}},\ }\href {\doibase 10.1103/RevModPhys.64.649} {\bibfield
  {journal} {\bibinfo  {journal} {Rev. Mod. Phys.}\ }\textbf {\bibinfo {volume}
  {64}},\ \bibinfo {pages} {649} (\bibinfo {year} {1992})}\BibitemShut
  {NoStop}%
\bibitem [{\citenamefont {{Cook}}\ \emph {et~al.}(1994)\citenamefont {{Cook}},
  \citenamefont {{Shapiro}},\ and\ \citenamefont {{Teukolsky}}}]{rotatingstar}%
  \BibitemOpen
  \bibfield  {author} {\bibinfo {author} {\bibfnamefont {G.~B.}\ \bibnamefont
  {{Cook}}}, \bibinfo {author} {\bibfnamefont {S.~L.}\ \bibnamefont
  {{Shapiro}}}, \ and\ \bibinfo {author} {\bibfnamefont {S.~A.}\ \bibnamefont
  {{Teukolsky}}},\ }\href {\doibase 10.1086/173934} {\bibfield  {journal}
  {\bibinfo  {journal} {\apj}\ }\textbf {\bibinfo {volume} {424}},\ \bibinfo
  {pages} {823} (\bibinfo {year} {1994})}\BibitemShut {NoStop}%
\end{thebibliography}%
\end{document}